\documentclass[twocolumn,prl,aps,longbibliography,superscriptaddress]{revtex4-2}
\usepackage[english]{babel}
\usepackage{amsmath}
\usepackage{mathtools}
\usepackage{graphicx}
\usepackage[colorlinks=true, allcolors=blue]{hyperref}
\usepackage{bm}
\usepackage{dsfont}
\usepackage{xcolor}
\usepackage{orcidlink}

\definecolor{myblue}{rgb}{0, 0.2, 1.0}

\begin{document}

\title{Non-local edge mode hybridization in the long-range interacting Kitaev chain}
    \author{David Haink\orcidlink{0009-0004-1573-6034}}
    \email{david.haink@dlr.de}
    \affiliation{High-performance Computing, Institute of Software Technology, German Aerospace Center (DLR), 51147 Cologne, Germany}
    \affiliation{Department of Physics, TU Dortmund University, 44227 Dortmund, Germany}
    
    \author{Andreas A. Buchheit\orcidlink{0000-0003-4004-713X}}
    \email{andreas.buchheit@uni-saarland.de}
    \affiliation{Department of Mathematics, Saarland University, 66123 Saarbr\"ucken, Germany}
    \affiliation{Department of Mathematics, ETH Zürich,  Rämistrasse 101, 8092 Zürich, Switzerland}
    \author{Christof Weitenberg\orcidlink{0000-0001-9301-2067}}
    \email{christof.weitenberg@tu-dortmund.de}
    \affiliation{Department of Physics, TU Dortmund University, 44227 Dortmund, Germany}
    \author{Benedikt Fauseweh\orcidlink{0000-0002-4861-7101}}
    \email{benedikt.fauseweh@tu-dortmund.de}
    \affiliation{High-performance Computing, Institute of Software Technology, German Aerospace Center (DLR), 51147 Cologne, Germany}
    \affiliation{Department of Physics, TU Dortmund University, 44227 Dortmund, Germany}

    \date{\textrm{\today}}

\begin{abstract}
In one-dimensional $p$-wave superconductors with short-range interactions, topologically protected Majorana modes emerge, whose mass decays exponentially with system size, as first shown by Kitaev. In this work, we extend this prototypical model by including power law long-range interactions within a self-consistent framework, leading to the self-consistent long-range Kitaev chain (seco-LRKC). In this model, the gap matrix acquires a rich structure where short-range superconducting correlations coexist with long-range correlations that are exponentially localized at both chain edges simultaneously. As a direct consequence, the topological edge modes hybridize even if their wavefunction overlap vanishes, and the edge mode mass inherits the asymptotic scaling of the interaction. In contrast to models with imposed power law pairing, where massive Dirac modes emerge for exponents $\nu < d$, we analytically motivate and numerically demonstrate that, in the fully self-consistent model, algebraic edge mode decay with system size persists for all interaction exponents $\nu > 0$, despite exponential wave function localization. While the edge mode remains massless in the thermodynamic limit, finite-size corrections can be experimentally relevant in mesoscopic systems with effective long-range interactions that decay sufficiently slowly. Our findings have direct implications for quantum simulations of ultracold microwave-shielded dipolar molecules in one-dimensional optical lattices.
\end{abstract}

\maketitle
\paragraph{Introduction}
Topological superconductors have garnered significant interest due to their potential applications in fault-tolerant quantum computing \cite{Kitaev2003}. These materials support edge modes that follow non-Abelian statistics, allowing to encode and manipulate quantum information using braiding operations. Since the information is encoded non-locally, this approach offers intrinsic error protection, as  correlated decoherence processes are suppressed for large spatial separation between the edge modes. As a result, topological quantum computing offers a promising route towards scalable quantum computing \cite{nayak2008non}. 

One of the most fundamental models that exhibits topological edge modes is the Kitaev chain \cite{Kitaev2001}. It supports Majorana zero modes (MZMs), zero-energy, non-Abelian quasiparticles that are their own antiparticles and are localized at the chain ends. Over the past two decades, considerable effort has been devoted to identify such MZMs in heterostructures theoretically \cite{Fu2008,PhysRevLett.104.040502,Lutchyn2010,Oreg2010,PhysRevB.84.144522,Sau2012,Pientka2013,PhysRevLett.129.267701} and experimentally \cite{Mourik2012,Das2012,Rokhinson2012,Deng2012,Churchill2013,NadjPerge2014,Nichele2017,Suominen2017,Ren2019,Aghaee2023}. 

Recently, significant advances have been made in realizing quantum-dot-based Kitaev chains \cite{tenHaaf2025,TenHaaf2024,Dvir2023,PhysRevX.13.031031,PhysRevLett.132.056602} as well as using quantum dots for parity readout \cite{Aghaee2025}.
In these systems, long-range interactions can arise naturally, including inter-dot Coulomb interactions \cite{PhysRevB.109.035415}, RKKY interactions mediated by the superconducting substrate \cite{Pientka2013}, and fluctuation-induced long-range interactions near a quantum critical point \cite{PhysRevB.96.134517}. In effective bilinear models, long-range interactions are often assumed to directly manifest as long-range pairing terms, where the superconducting gap of the Kitaev chain is parametrized as a power law $\Delta_{x,x'}\sim |x-x'|^{-\nu}$, where $x, x'$ are the lattice sites of the paired electrons and $\nu$ the power law exponent. We refer to models of this kind as non-self-consistent long-range Kitaev chains (non-seco-LRKC). Depending on the power law exponent, this type of pairing can significantly alter the phase diagram of the Kitaev chain \cite{Viyuela2016,Alecce2017}. 

The physics of non-seco-LRKCs has been extensively studied \cite{Vodola2014,Bhattacharya2018,Francica2023, Jones2023,Lepori2017a,Lepori2018,Lepori2023,Ortiz2014,Patrick2017,Vodola,Yu2020}, revealing that in the thermodynamic limit MZMs persist for weak long-range pairing characterized by an algebraic decay with power law exponent $\nu>d$, where we here focus solely on $d=1$ systems. In contrast, strong long-range pairing ($\nu<d$) results in massive Dirac fermions. Previous studies suggest that the transition has continuously-varying critical exponents that depend both on $\nu$ and on the system dimension \cite{Cai2017,Ansari2021,Baghran2024,Bhattacharya2019,Cats2018,Defenu2019,Kartik2021,King2021,Lepori2016,Maity2019,Minato2022,Mitra2025,Ren2022,Yang2022,Zerba2023,Niu2025}. Additionally, entanglement entropy investigations show that universality breaks down at critical power laws \cite{Ercolessia,Ares2018,Ares2019,Francica2022,Lepori2022,Lerose2020,Ma2024,Mitra2024,Mondal2022,Nandy2018,Pezze2022,Pezze2017}.
Long-range pairing has a profound impact on the non-equilibrium properties of the chains. Current transport phenomena have been explored in terms of Andreev states within the Kitaev ladder framework \cite{Banerjee2023,Giuliano2018,Liu2020,Nehra2020,Ren2021,banerjee2025transport}, showing reduced localization due to long-range effects. Non-equilibrium dynamics \cite{Uhrich2020,VanRegemortel2016}, the dynamics following quenches \cite{Canoy2018,Defenu2019a,Dutta2017,Huang2024,Starchl2024,Su2020,Vodola2015a,Lepori2017} and finite temperature effects \cite{Cinnirella2024,HernandezSantana2017,Jin2024,King2022,Solfanelli2023a,Wang2020} show qualitatively different features than in the short range case.
A comprehensive analysis of edge modes has demonstrated that localized states decay according to the power law $\nu$ \cite{Botzung2019,Jaeger2020,Radgohar2020,Tarantola2023,Zhong2024}. Similar investigations have been conducted for models such as the Aubry-André-Harper model \cite{Fraxanet2021,Fraxanet2022,Gandhi2024} and Shiba chains \cite{Pientka2015}. 

In all the non-seco-LRKC models discussed above, long-range pairing appears as a bilinear term in the Hamiltonian. Although the terms \emph{long-range interaction} and \emph{long-range pairing} are often used interchangeably, it is important to distinguish between them. The extent to which the pairing inherits the spatial dependence of the underlying interaction has, so far, remained poorly understood, as does the resulting impact on topology and edge modes.

In this work, rather than imposing long-range pairing directly, we adopt a more general approach by introducing an effective long-range density-density interaction that decays algebraically with exponent $\nu$. Applying BCS theory within a standard mean-field framework, we derive a self-consistent long-range Kitaev chain (seco-LRKC), in which the pairing emerges naturally from the underlying interaction. We then investigate the qualitative differences between non-self-consistent (non-seco) and self-consistent (seco) LRKCs, with particular focus on edge mode properties, such as their mass and their finite-size scaling behavior with respect to chain length. Finally, we argue that the results of this work should become experimentally accessible in near-term experiments with microwave-dressed polar fermionic molecules in one-dimensional optical lattices.

\paragraph{A long-range self-consistent Kitaev chain}
We start with a spinless or spin-polarized BCS superconducting Hamiltonian $H=H_0+H_{\text{int}}$,
where $H_0=  -\frac{1}{2}\sum_{x=1}^{n-1}\tau c^\dag_{x}c_{x+1}+ \mathrm{h.c.} -\frac{1}{2}\sum_{x=1}^{n}\mu (c^\dag_{x}c_{x}-\frac{1}{2})$ and $H_{\text{int}}=\frac{1}{2}\sum_{x\neq x'}c^\dag_{x}c^\dag_{x'}V_{x,x'}c_{x'}c_{x}$
are the hopping and interaction Hamiltonian respectively. Here, $\tau > 0$ is the hopping amplitude, $c^{(\dagger)}_x$ denotes the annihilation (creation) operator for an electron at site $x$ in a one-dimensional lattice of $n$ sites, and $\mu$ is the chemical potential. The interaction between electrons located at different sites $x$ and $x'$ is given by $V_{x,x'}$. We assume that $V$ results from a renormalized description and can be described by the following attractive long-range power-law
\begin{equation}
    V_{x,x'}=
		-U_0|x-x'|^{-\nu},
\end{equation} 
where $U_0>0$ indicates the interaction strength and $\nu$ denotes the long-range power law exponent \cite{Buchheit2023}.
A standard mean-field approximation
\begin{equation}
	\big(c^\dag_{x}c^\dag_{x'}-\langle c^\dag_{x}c^\dag_{x'}\rangle\big)\big(c_{x'}^{\phantom{\dag}} c_{x}^{\phantom{\dag}} - \langle c_{x'}^{\phantom{\dag}} c_{x}^{\phantom{\dag}} \rangle\big)\approx0,
\end{equation}
then yields the bilinear mean-field interaction Hamiltonian
\begin{equation}
\label{Eq_int_ham}
    H_{\text{int}}\approx\frac{1}{2}\sum_{x,x'}\left[  (\Delta_{x,x'} c^\dag_{x}c^\dag_{x'}+\mathrm{h.c.}) 
     +\Delta_{x,x'} \langle c^\dag_{x}c^\dag_{x'}\rangle\right],
\end{equation}
with the real-space superconducting gap matrix 
$
    \Delta_{x,x'} = V_{x,x'}\langle c_{x'}c_{x}\rangle.
$
We briefly comment on the validity of this self-consistent mean-field description.
First, sufficiently long-ranged interactions are known to suppress fluctuations and render mean-field theory asymptotically exact \cite{Defenu2023}, which is often interpreted as an effective increase in system dimension. Second, in quantum lattice systems with sufficiently long-ranged interactions, the critical exponents assume their mean-field values \cite{Adelhardt2025}.
Finally, we confirm, following \cite{Ortiz2014}, the topological nature of the fully interacting model in the supplementary materials using exact diagonalization of small chains. Note that, due to long-range interactions, spontaneous symmetry breaking is not forbidden in one-dimension in the thermodynamic limit as the Mermin-Wagner theorem does not apply \cite{Maghrebi2017}.
The standard non-seco case is obtained by imposing a gap such that $\Delta_{x,x'}= \mathrm{sgn}(x-x') \Delta_\text{const}|x-x'|^{-\nu}$, where $\Delta_\text{const}$ is a complex constant. In contrast to that, the gap solution of the seco-LRKC is obtained as follows. First, the Hamiltonian is rewritten in particle-hole symmetric Nambu-spinor form $H=\frac{1}{2}\bm{\Psi}^\dag\mathcal{H}\bm{\Psi}+\mathrm{const}$, with $\bm{\Psi}=(c_{1},c_{2},...,c^\dag_{1},c^\dag_{2},...)^T$ and  with the Bogoliubov-de-Gennes (BdG) matrix
\begin{equation}
	\mathcal H=\left( 
	\begin{matrix}
		h/2&\Delta\\
		\Delta^\dagger&-h/2\\
	\end{matrix}
	\right).
    \label{eq:BdGmatrix}
\end{equation}
The hopping matrix $h\in \mathds R^{n\times n}$ is defined as $h_{x,x}=-\mu$ and $h_{x,x\pm 1}=-\tau$ and zero otherwise. 
%Here we have used the fermionic anticommutation relations $\{c_x,c_{x'}\}=\{c^\dag_x,c^\dag_{x'}\}=0,\{c^\dag_x,c_{x'}\}=\delta_{x,x'}$.
The skew symmetric correlation matrix $\alpha_{x,x'}=\langle c_{x'} c_x \rangle$ of the seco-LRKC can then be readily determined (up to a global $U(1)$ phase) as the minimum of the nonlinear energy functional 
\begin{equation}
E[\alpha]=-\frac{1}{2}\sum_{x\neq x'}V_{x,x'}|\alpha_{x,x'}|^2-\frac{1}{4}\operatorname{tr}\sqrt{\mathcal{H}^2}.\label{eq:energy_functional}
\end{equation} The BdG matrix in Eq.\,\eqref{eq:BdGmatrix} and the self-consistent gap obtained from minimizing the energy in Eq.\,\eqref{eq:energy_functional} then define the seco-LRKC for general attractive interactions $V$. 

\paragraph{Structure of self-consistent gap solution}

\begin{figure*}
\centering
\includegraphics[width=1.0\textwidth]{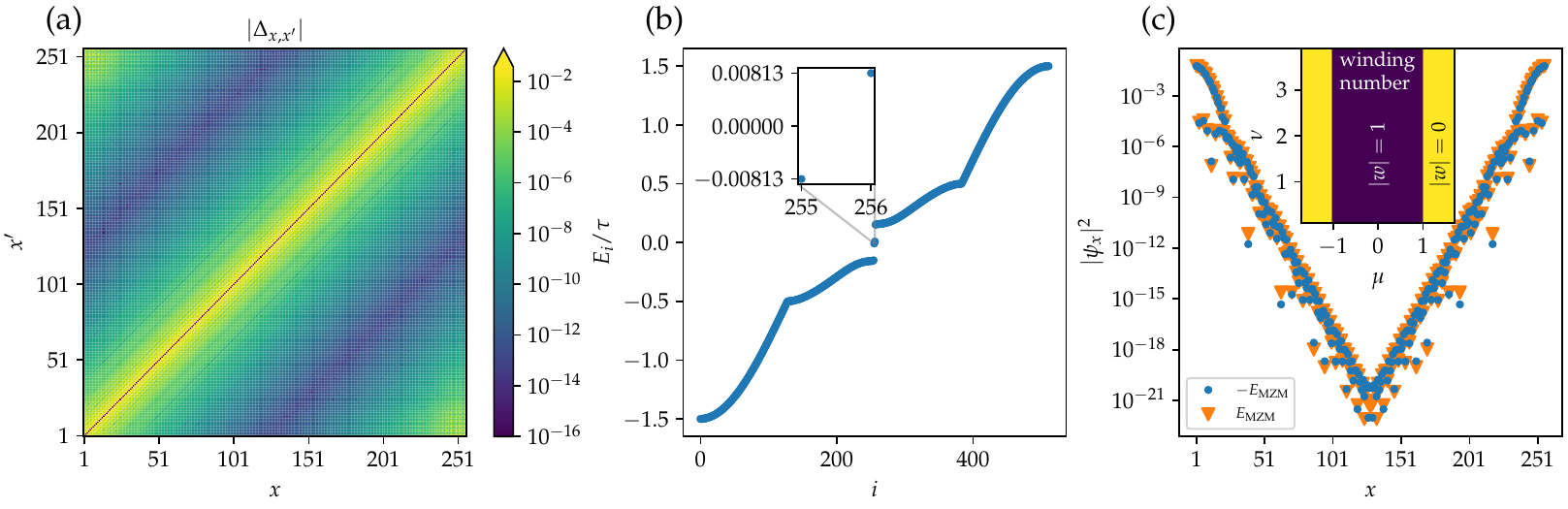}
\caption{\label{fig:eig_slr}Self-consistent solution for $U_0=\tau/2,\ \nu=1/2,\ \mu=\tau/2,\ n=256$. (a) Self-consistent gap $\Delta_{x,x'}$ as a function of $x$ and $x'$. (b) Eigenvalues of the BdG matrix $\mathcal H$ with self-consistent gap. The inset shows the separation of the edge modes. The bulk is gapped and its largest value is $E=\tau+|\mu|$. For $E=\pm |\mu|$ we observe a Van-Hove singularity. (c) Eigenstates of both coupled edge modes, where only the annihilation operator part, corresponding to the first $n$ entries, of $\bm \Psi$ is shown. The inset shows the phase diagram as a function of chemical potential $\mu$ and long-range exponent $\nu$. Yellow indicates a winding number of $w=0$, while violet denotes $|w|=1$.}
\end{figure*}
 
In the following, we determine the self-consistent gap from Eq.~\eqref{eq:energy_functional} for a finite chain with $n$ lattice sites. For $|\mu| < \tau$, a p-wave solution emerges. The resulting absolute value of self-consistent gap $\Delta_{x,x'}$ is shown for the parameter values $U_0 = \tau/2$, $\mu = \tau/2$, and $\nu = 1/2$ in Fig.\,\ref{fig:eig_slr}(a). For general parameter choices with $\nu > 0$ and $0 < U_0 < 2\tau$, we find that the corresponding correlation matrix $\alpha_{x,x'}$ always exhibits the same qualitative structure. For sufficiently large system sizes $n$, the correlation matrix separates into two disjunct contributions $\alpha=\alpha^{\mathrm{sr}}+\alpha^\mathrm{lr}$, with short-range correlations $|\alpha^{\mathrm{sr}}_{x,x'}|\sim g_1 e^{-|x-x'|/\lambda_1}$ exponentially localized around the main diagonal and $|\alpha^\mathrm{lr}_{x,x'}|\sim g_2 e^{-(|x-x'|-n)/\lambda_2}$ exponentially localized at the anti-diagonal corners of the correlation matrix $x=1$, $x'=n$, and vice versa. The weights $g_1,g_2>0$ and widths $\lambda_1,\lambda_2$ further converge to constant non-zero values for $n\to \infty$ and we numerically observe $\lambda_1 \approx \lambda_2$. The short-range band corresponds to the translationally invariant bulk solution for $1\ll x+x'\ll 2n$ with $\alpha^{\mathrm{sr}}_{x,x'}=\alpha^\mathrm{bulk}_{x-x'}$, where  $\alpha^\mathrm{bulk}$ can also be recovered by solving the gap equation in Fourier space, see i.e.~\cite{Buchheit2023}. Finite-size corrections to the bulk solution are exponentially localized at the chain edges.
 
In conclusion, the correlation matrix $\alpha$ separates into a short-range band localized around the main diagonal and, importantly, exponentially localized long-range correlations at the anti-diagonal corners, all of which converge to a sparse matrix with constant substructures as $n\to \infty$. This non-trivial band structure of the correlation matrix, that naturally emerges when the correlations originate self-consistently due to interactions, is in strong contrast with an externally imposed superconducting correlations in non-seco models, where $\alpha_{x,x'}\propto\mathrm{sgn}(x-x')$ is dense and effectively constant. 
The above band structure of the correlations is inherited by the gap matrix. Here, the band around the main diagonal is influenced solely by the near-field part of the interaction $V_{x,x'}$ where $|x-x'|<\lambda_1$. Intermediate-range interactions at distances $\lambda_1\ll |x-x'|\ll n$ do not enter at all in the gap, as the associated correlations vanish. Finally, the far long-range tail of the interaction scaling as $n^{-\nu}$ is encoded in the exponentially localized long-range blocks $\Delta^\mathrm{lr}$ of the gap matrix. This cluster has important consequences for the qualitative behavior of the edge modes. A posteriori, this structure can be understood by treating the long-range interaction as a perturbation of the short-range Kitaev chain, where long-range correlations associated with the edge modes are known to occur \cite{miao2018majorana}. Through the self-consistency relation $\Delta_{x,x’}=V_{x,x’}\alpha_{x,x’}$, these correlations enter directly into the long-range contribution $\Delta^{\mathrm{lr}}$ of the gap.

\paragraph{Non-local edge mode hybridization}

\begin{figure*}
\centering
\includegraphics[width=1.0\textwidth]{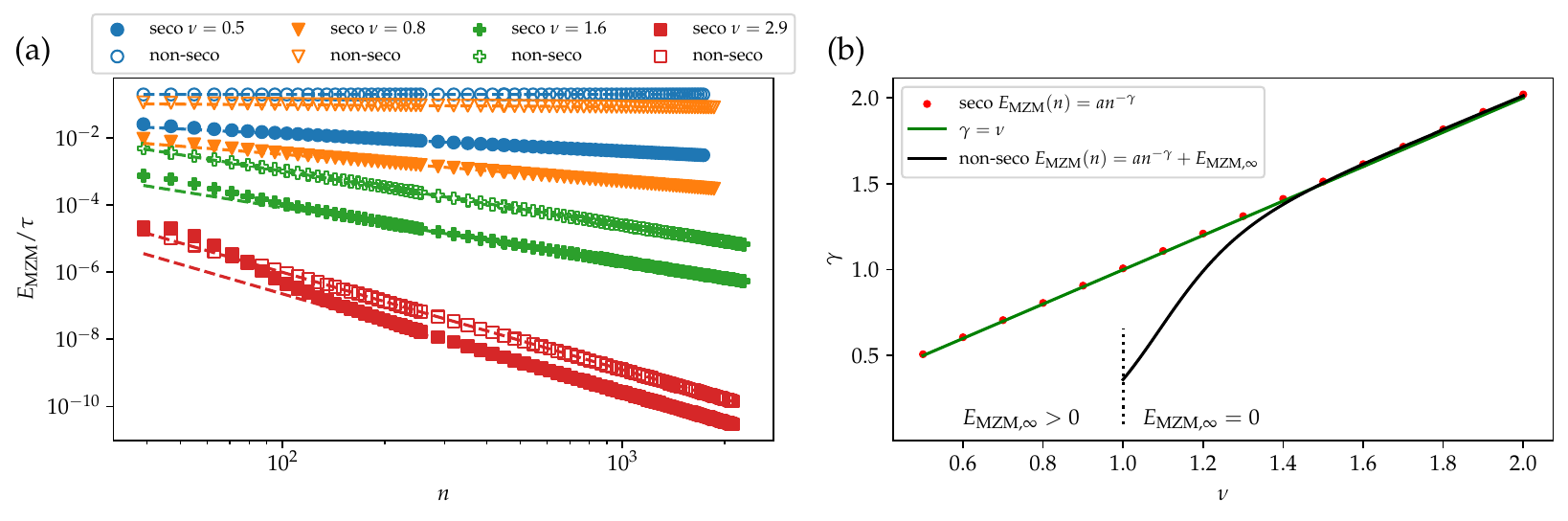}
\caption{\label{fig:fse} (a) Absolute value of the edge mode eigenvalue as a function of system size $n$ for different interaction exponents $\nu$, contrasting self-consistent and non-self consistent solutions with parameters $U_0=\tau/2$ and $\mu=0$.  The dashed lines represent a power law fit to the data. (b) Decay exponent $\gamma$ of the edge mode eigenvalue (obtained by the fit in (a)) as a function of the interaction exponent $\nu$. $E_{\mathrm{MZM},\infty}$ is a fitting parameter for the mass in the thermodynamic limit.
}
\end{figure*}

\begin{figure}
%\centering
\includegraphics[width=0.5\textwidth]{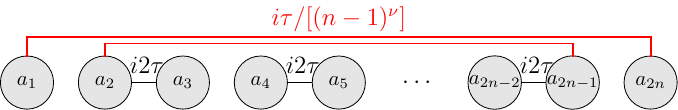}
\caption{\label{fig:tikz} Edge mode hybridization in the picture of Majorana operators for $U_0=2\tau$.}
\end{figure}

Having identified the emergent gap structure in the seco-LRKC, we now analyze its consequences for the properties of the MZMs, as they become massless as $n\to \infty$,  while contrasting them with the non-self-consistent model. The edge modes are identified by the eigenvectors of the BdG matrix corresponding to the lowest absolute eigenvalues. We depict the band structure of $\mathcal{H}$ in Fig.\,\ref{fig:eig_slr}(b). The corresponding edge mode eigenvectors (within the creation channel) are shown in panel (c). They form superposition states simultaneously localized at both ends of the chain, with wave functions that decay exponentially into the bulk from either end on the scale $\lambda_2$.
Hybridization of MZMs due to finite wave-function overlap is an important effect in topological qubit devices. Such hybridization breaks the ground-state degeneracy and induces a finite MZM mass $E_\mathrm{MZM}$, i.e. a finite energy splitting of the otherwise zero-energy edge modes, which in turn leads to unintended qubit dephasing $\sim e^{-iE_\mathrm{MZM} t}$ during braiding operations, see, e.g.,~\cite{hodge2025characterizing}. Although the edge modes in our model are exponentially separated (see Fig.\,\ref{fig:eig_slr}(c)), they remain hybridized with a finite MZM mass. We refer to this effect as non-local edge mode hybridization. 
The origin of the non-local hybridization lies in the $\Delta^\mathrm{lr}$ clusters of the gap matrix. If these clusters are artificially removed, the edge mode energy decays exponentially with system size $n$, similar to the standard Kitaev chain. This behavior follows from the topologically nontrivial $p$-wave pairing combined with the exponential decay of finite-size corrections of the bulk contribution. However, since the gap clusters scale with the power law tail of the interaction $\lvert \Delta^\mathrm{lr} \rvert \sim n^{-\nu}$, first-order perturbation theory suggests that the Majorana zero-mode energy inherits the scaling
\begin{equation}
|E_\mathrm{MZM}| \propto n^{-\nu},
\label{eq:asymptotic_scaling}
\end{equation}
for arbitrary exponents $\nu$. Thus, in the limit $n \to \infty$, the edge mode mass vanishes and the modes decouple. For finite mesoscopic systems with power law interaction tails, however, a residual mass remains, that decays only algebraically with chain length. In Fig.~\ref{fig:fse}\,(a) we present the edge mode mass for various exponents $\nu$ as a function of $n$, observing a power law scaling in the self-consistent model. Panel (b) shows the corresponding decay exponent $\gamma$, extracted from a fit to $E_\mathrm{MZM}(n)=an^{-\gamma}$, confirming the prediction $\gamma = \nu$ from Eq.~\eqref{eq:asymptotic_scaling}.
To understand the physical origin of edge-mode hybridization, we rewrite the Hamiltonian in terms of the Majorana operators introduced in \cite{Kitaev2001}. We find in the seco-LRKC at $\mu=0$ and in the limit $U_0\to 2\tau$ that the short-range part of the correlation matrix becomes exactly localized to nearest neighbors, so that the bulk Hamiltonian reduces to the nearest-neighbor form of the short-range Kitaev chain.
The Hamiltonian then reads $H=\frac{i\tau}{2}\sum_{j=1}^{n-1}a_{2j}a_{2j+1}+\frac{i\tau}{4(n-1)^\nu}[a_1a_{2n}+a_2a_{2n-1}]$, where $a_{2j}=(c_j-c_j^\dag)/i, a_{2j-1}=c_j+c_j^\dag$ are the Majorana operators. 
The edge Majorana operators $a_1$ and $a_{2n}$ appear only in the  bilinear term $\frac{i\tau}{4(n-1)^\nu}a_1a_{2n}$, which is the microscopic origin of the non-local edge-mode hybridization. Similarly, $a_2$ and $a_{2n-1}$ are coupled via the same long-range term and, as illustrated in Fig.~\ref{fig:tikz}, remain connected to the bulk via their nearest-neighbor couplings to $a_3$ and $a_{2n-2}$, respectively. This yields a small correction to the bulk energy $\tau/2$ for these modes \cite{Supplement}. Although this explicit Majorana form is obtained in a particular limit, the same mechanism persists for general gap matrices where the bulk spectrum exhibits a gap. 
We verified that the non-local edge mode hybridization is robust against disorder in the chemical potential and in the hopping and also against dimerization in the hopping.

Let us now compare these results with the non-seco model, where $\Delta_{x,x'} \sim \mathrm{sign}(x-x') V_{x,x'}$. 
Here we observe convergence to a constant $E_{\text{MZM}} > 0$ for long-range exponents $\nu < 1$, obtaining massive Dirac fermions, i.e. a finite energy splitting of the edge mode energies that persists for infinite system sizes, see Ref.~\cite{Viyuela2016}. Even for $\nu > 1$, differences in the scaling behavior of the edge mode masses arise. While the seco model strictly inherits the interaction exponent $\nu$ (red dots on green line in Fig.~\ref{fig:fse}(b)), the non-seco model exhibits a smaller decay exponent $\gamma$ (black). The scaling behavior of the two models coincides approximately for $\nu >3/2$.
\paragraph{Phase diagram of the long-range BCS Kitaev chain}
We now investigate the topology of the seco-LRKC by computing the bulk winding number following Ref.~\cite{Lin2021}. Here, we find that the phase diagram simplifies compared to the non-self-consistent model. As shown in the inset of Fig.~\ref{fig:eig_slr}(c), the winding number is independent of the power law exponent $\nu$. As a function of the chemical potential $\mu$, we find only the trivial transition at $|\mu| = \tau$, where superconductivity vanishes, recovering the result of the standard Kitaev chain.
This behavior contrasts with the non-seco case, which exhibits a second topological phase for $\nu < 1$ with massive Dirac fermions, as well as a transition to the region $1 < \nu < 3/2$ where finite-size scaling no longer matches the pairing exponent (see Fig.~\ref{fig:fse}(b)). Moreover, unlike the non-seco model \cite{Viyuela2016}, the seco solution shows no continuation of the topological phase for $\mu < -1$ when $\nu < 3/2$.
\paragraph{Conclusion and Outlook}
In this work, we investigated the self-consistent long-range Kitaev chain (seco-LRKC), an extension of the original Kitaev chain in which the superconducting gap $\Delta_{x,x'}$ arises self-consistently from an attractive electron–electron interaction with a long-range tail. We have highlighted key qualitative differences from the well-studied non-self-consistent (non-seco) models, where long-range pairing is assumed. In the self-consistent model the superconducting gap develops a sparse band structure, which significantly affects the properties of the Majorana zero modes (MZMs). This structure persists for generic parameter choices within the superconducting phase.
We demonstrated that short-range pairings are exponentially suppressed with distance $x-x'$. 

The width of these pairings determines the spatial support of the edge mode wavefunction. Additionally, an exponentially localized long-range cluster $\Delta^\mathrm{lr}$ emerges at the ends of the anti-diagonal of the gap matrix (localized at $x=1, x'=n$ and vice versa), which inherits the scaling $\sim n^{-\nu}$ from the interaction tail. 
This edge cluster couples the two edge modes, even when the edge mode wavefunctions do not overlap, leading to non-local edge mode hybridization, lifting the ground-state degeneracy. 
The resulting edge mode mass decays algebraically with system size $E_\mathrm{MZM}\sim n^{-\nu}$  for any exponent $\nu > 0$, yielding true zero modes in the thermodynamic limit $n \to \infty$. In the self-consistent model, the phase diagram simplifies, leaving only two distinct phases: the trivial and topological phases.% as in the original Kitaev chain. 
This behavior is in sharp contrast to non-seco models, which feature a dense gap matrix $|\Delta_{x,x'}| \propto |V_{x,x'}|$, a resulting power law scaling of the edge mode wavefunction \cite{Botzung2019}, and a transition to massive Dirac fermions for $\nu < 1$, along with other changes to the phase diagram.

Our results may have direct experimental relevance for the search for MZMs in systems where electron–electron interactions possess long-range tails.
Specifically, we demonstrate in the supplemental material that microwave-dressed dipolar fermionic molecules \cite{PhysRevLett.103.155302} confined to a one-dimensional geometry in an optical lattice realize our model with an effectively attractive $1/r^{3}$ interaction.  Recent experiments have created degenerate gases \cite{Schindewolf2022,Bigagli2024} and reached a strongly-interacting regime \cite{zhang2025} and it is to be expected that further improvements will allow reaching temperatures down to $0.1 T_F$ \cite{Wang2024}. This temperature regime will allow probing the edge states discussed in this article. The decay of the superconducting order parameter could be directly probed in a system of ultracold molecules by implementing a many-body phase microscope \cite{weitenberg2026}.

Other potential systems are  quasi-1d p-wave superconductors such as K$_2$Cr$_3$As$_3$ \cite{PhysRevX.5.011013,doi:10.1126/sciadv.abl4432}, where magnetic spin fluctuations can mediate triplet pairing \cite{PhysRevLett.115.227001}, if Coulomb screening is incomplete and interactions acquire long-range tails. More generally, the mechanism described in this work is relevant to edge modes in any mesoscopic structure, where superconductivity arises from effective interactions that extend across the device scale.

In these cases, superconductors may host finite-energy edge modes that remain topological but acquire a mass decaying only algebraically with system size. In mesoscopic systems, this decay can be too slow for the edge modes to reach true zero energy, leading to unintended qubit dephasing during braiding operations. Our work thus points to an important dephasing mechanism that would be overlooked in the conventional non-self-consistent model.

It will be interesting to investigate also other properties of the long-range interacting topological superconductors in higher dimensions \cite{buchheit2025computation}, where zero modes occur in vortex cores. We expect similar differences to the non-self-consistent chain for quantities, such as the entanglement entropy \cite{Ercolessia,Ares2018,Ares2019,Francica2022,Lepori2022,Lerose2020,Ma2024,Mitra2024,Mondal2022,Nandy2018,Pezze2022,Pezze2017}, current transport phenomena \cite{Banerjee2023,Giuliano2018,Liu2020,Nehra2020,Ren2021,banerjee2025transport}, information exchange out of equilibrium \cite{Uhrich2020,VanRegemortel2016}, finite temperature effects \cite{Cinnirella2024,HernandezSantana2017,Jin2024,King2022,Solfanelli2023a,Wang2020} and the dynamics following quenches \cite{Canoy2018,Defenu2019a,Dutta2017,Huang2024,Starchl2024,Su2020,Vodola2015a,Lepori2017}. 

\section*{Acknowledgements}
The authors thank Jonathan Busse and Torsten Keßler for fruitful discussions. We also thank Daniel Seibel for providing us with his problem-tailored L-BFGS algorithm. A.B. is grateful for the support and hospitality of the Pauli Center for Theoretical Studies at ETH Zürich. The data and code produced for this work are available at \cite{haink_2025_17277639}. This work was supported by the Klaus-Tschira Stiftung under Grant No. 00.025.2025.

\bibliography{cites.bib}
\end{document}